\documentclass[11pt,a4paper]{article}

\usepackage{graphicx}% Include figure files
\usepackage{epsfig}

\usepackage{hyperref}
\usepackage{graphicx,epsf}
\usepackage{bm}
\usepackage{amsmath}

\begin{document}

\begin{center}
{\large{\textbf{Shear Alfv\'en wave continuous spectrum within magnetic islands}}}\\
\vspace{0.4 cm}
{\normalsize {A. Biancalani${}^{1,2}$, L. Chen${}^{3,4}$, F. Pegoraro${}^5$, F. Zonca${}^{6}$}\\}
\vspace{0.2 cm}
\small{${}^1$ Max-Planck-Institut f\"ur Plasmaphysik, Euratom Association, D-85748 Garching, Germany\\
${}^2$ Max-Planck-Institut f\"ur Sonnensystemforschung, Katlenburg-Lindau, Germany\\
${}^3$ Institute for Fusion Theory and Simulation, Zhejiang Univ., Hangzhou, People's Rep. of China\\
${}^4$ Department of Physics and Astronomy, University of California, Irvine, CA 92697-4575, USA\\
${}^5$ Department of Physics, University of Pisa, 56127 Pisa, Italy\\
${}^6$ Associazione Euratom-ENEA sulla Fusione, C.R. Frascati, C.P. 65 - 00044 Frascati, Italy}
\end{center}

\begin{abstract}
The radial structure of the continuous spectrum of shear Alfv\'en waves is calculated in this paper within the separatrix of a magnetic island.
Geometrical effects due to the noncircularity of the flux surface's cross section are retained to all orders. On the other hand, we keep only curvature effects responsible for the beta-induced gap in the low-frequency part of the continuous spectrum.
Modes with different helicity from that of the magnetic island are considered.
The main result is that, inside a magnetic island, there is a continuous spectrum very similar to that of tokamak plasmas, where a generalized safety factor $q$ can be defined
and where a wide frequency gap is formed, analogous to the ellipticity induced Alfv\'en eigenmode gap in tokamaks. The presence of this gap is due to the strong eccentricity of the island cross section.
The importance of the existence of such a gap is recognized in potentially hosting magnetic-island induced Alfv\'en eigenmodes (MiAE).
Due to the frequency dependence of the shear Alfv\'en wave continuum on the magnetic-island size, the possibility of utilizing MiAE frequency scalings as a novel magnetic-island diagnostic is also discussed.
\end{abstract}


\section{Introduction}
\label{sec:intro}
Shear Alfv\'en waves (SAW) are electromagnetic plasma waves propagating as transverse waves along the magnetic field, with the characteristic Alfv\'en velocity $v_A=B/\sqrt{4\pi\varrho}$ ($B$ is the magnetic field and $\varrho$ the mass density of the plasma). In fusion plasmas, fast ions in the MeV energy range have velocities comparable with the typical Alfv\'en speed. In addition, SAW group velocity is directed along the magnetic field line and, therefore, fast ions can stay in resonance and effectively exchange energy with the wave~\cite{zonca06,chen07}. SAW in a nonuniform equilibrium experience collisionless dissipation (\emph{continuum damping}~\cite{Grad69,hasegawa74,chen74}), due to singular structures that are formed where the SAW continuum is resonantly excited. In tokamaks, the magnetic field intensity varies along the field line. This creates gaps in the SAW continuous spectrum~\cite{kieras82} due to translational symmetry breaking, analogous to electrons traveling in a periodic lattice~\cite{zonca06,chen07}.

Two types of collective shear Alfv\'en instabilities exist in tokamak plasmas: energetic-particle continuum modes (EPM)~\cite{chen94}, with frequency determined by fast particle characteristic motions, and discrete Alfv\'en eigenmodes (AE), with a frequency inside SAW continuum gaps~\cite{cheng85}. The former can become unstable provided the drive exceeds a threshold determined by the continuum damping absorption; the latter, have a generally lower instability threshold, being practically unaffected by continuum damping~\cite{zonca06,chen07,Grad69,hasegawa74,chen74}. For this reason, the importance of understanding the continuum radial structure is clear, if one faces the problem of reaching the ignition condition with magnetically confined fusion plasmas, due to the potential impact of AE stability on the plasma confinement.

The nonlinear dynamics of SAW is a research topic still rich of open issues.
In the special case of uniform ideal incompressible plasmas a peculiar state exists, called the \emph{Alfv\'enic state}~\cite{Alfven50,Elsasser56,Hasegawa89}, where the nonlinear effects due to Maxwell and Reynolds stresses cancel and give a self-consistent nonlinear state. On the other hand, in tokamak plasmas, nonuniformities and non-ideal effects such as resistivity or finite plasma compressibility are present. This breaks the Alfv\'enic state conditions and therefore SAW are characterized by a rich nonlinear dynamics. As a consequence, the SAW continuous spectrum can be modified by the interaction with low-frequency MHD fluctuations~\cite{buratti05nufu}, such as magnetic islands. Magnetic islands are the result of magnetic-field reconnection processes, which occur in tokamak plasmas because of non ideal effects, such as finite resistivity~\cite{furth63}.
The typical island oscillation frequency and growth rate are much lower than the SAW frequency. Due to this time scale separation between island and the SAW dynamics, the island is modelled here as a static helical distortion of the equilibrium.

We derive our fluid theoretical description of the SAW continuum structure inside a finite size magnetic island in finite-$\beta$ tokamak plasmas. The magnetic island is modelled here as a straight flux tube with a noncircular cross section. Therefore, we neglect toroidicity effects on coupling modes with different poloidal mode numbers. However, we do take account the low-frequency gap in the SAW continuum due to  the effects of geodesic curvature and finite compressibility~\cite{chu92,turnbull93,zonca96}. In our model, the coupling of SAW and the island magnetic perturbation is a linear problem of SAW dynamics in a modified static equilibrium. The local differential equation yielding the SAW continuum structure is solved numerically with a shooting method code, in the spatial range of interest inside the magnetic island~\cite{Biancalani2010prl}. We are interested here in modes with different helicity from that of the magnetic island; in other words, we assume $\partial / \partial \zeta \ne 0$, where $\zeta$ is the coordinate along the flux tube. The solution of this problem inside and outside the magnetic island, for modes with the same helicity of the magnetic island, is given in a different paper~\cite{Biancalani2010ppcf}.

As suggested by the magnetic field line helicity behavior~\cite{swartz84,buratti05nufu}, in an equilibrium with a magnetic island a generalized safety factor $q$ can be defined for each flux surface, as an appropriate average of the ratio of the magnetic field components along and perpendicular to the magnetic axis.
This safety factor is a monotone function of the distance from the O point, and grows towards the separatrix, where it has a singularity. This is due to the vanishing of the magnetic field intensity near the X point of the magnetic island.
Within the separatrix, we find a continuous spectrum very similar to that of tokamak plasmas, where the coupling of modes with different mode numbers creates frequency gaps. In particular, a wide frequency gap is formed~\cite{Biancalani2010prl}, analogous to the ellipticity induced Alfv\'en Eigenmode~\cite{betti91} (EAE) gap in tokamaks.
The presence of this gap is due to the strong eccentricity of the island cross section.
In this gap, a magnetic-island induced Alfv\'en eigenmode (MiAE) could exist as a bound state, essentially free of continuum damping, and interact nonlinearly with the magnetic island,  
if the thermal or energetic component of the plasma provide sufficient free energy for driving the mode.
Due to the frequency dependence of the shear Alfv\'en wave continuum on the magnetic-island size, the possibility of utilizing MiAE frequency scalings as a novel magnetic island diagnostic is also discussed.

The scheme of the paper is the following. In Sec.~\ref{sec:equilibrium-coordinates} we describe the equilibrium magnetic field, starting with the tokamak coordinates and introducing the cylinderlike coordinates for the magnetic island flux tube. The generalized safety factor defined inside the magnetic island, is shown to have similar behavior of the safety factor in diverted tokamaks. In Sec.~\ref{sec:continuous-spectrum}, the SAW model equations are introduced and the equation for general shear Alfv\'en modes is provided. The equation for continuum modes is then derived in the cylinderlike coordinate set defined inside the island and the eigenvalue problem is established. The resulting continuous spectrum, obtained by solving numerically the eigenvalue problem with a shooting method code, is shown to be very similar to the continuous spectrum of tokamaks. The result is provided for typical tokamak parameters, and typical size magnetic islands. The analogy with the tokamak SAW continuum is shown in the limit of very large magnetic islands, when the section of the magnetic island flux tube is nearly circular. Finally Sec.~\ref{sec:Discussion} is devoted to a discussion about the applications of our results, in gaining a better knowledge on the stability of tokamak plasmas. The differential operators calculated in the cylinderlike coordinate set are provided in Appendix~\ref{appendix:differential-operators}.

\section{Equilibrium and coordinate system}
\label{sec:equilibrium-coordinates}

\subsection{Tokamak coordinates}

We consider a tokamak geometry where $R_0$ is the major radius of the torus. The equilibrium is made of an axisymmetric tokamak magnetic field, plus a helical perturbation generating the magnetic island. We start our analysis by describing the axisymmetric field with an orthogonal coordinate system $(r_T,\theta_T,\zeta_T)$ where $r_T$ is the radial coordinate, $\theta_T$ is the poloidal angle, corresponding to the field $\bm{B}_{pol} =B_{pol}(r_T) \, \bm{\hat{\theta}}_T$, and $\zeta_T$ the toroidal angle, corresponding to the toroidal field $\bm{B}_{tor}=B_{tor}\bm{\hat{\zeta}}_T$, considered uniform for simplicity. The subscript $T$ denotes here the tokamak coordinates. The gradients associated with these coordinates are: $\bm{\nabla} r_T = \hat{\bm{r}}_T$, $\bm{\nabla} \theta_T = \hat{\bm{\theta}}_T/r_T$ and $\bm{\nabla} \zeta_T = \hat{\bm{\zeta}}_T/R_0$.

We want to describe with a slab model, the region of the tokamak plasma where a magnetic island with poloidal and toroidal numbers $(m_{isl},n_{isl})$ is located. This is the region around the magnetic island's rational flux surface $q_T = q_0 = m_{isl}/n_{isl}$, corresponding to $r_T=r_0$, where the axisymmetric field has value $B_0 = \gamma B_{tor}$, with $\gamma = \sqrt{1+\varepsilon_0^2/q_0^2}$. Here $q_T$ is the tokamak safety factor $q_T = r_T B_{tor}/(R_0 B_{pol})$, and $\varepsilon_0 = r_0 /R_0$ is the inverse aspect ratio, assumed to be small: $\varepsilon_0 \ll 1$.
Therefore, we introduce an orthogonal coordinate system $(q_T,u,\zeta)$ through a rotation in the ($\theta_T,\zeta_T$) plane, defining $u=n_{isl}(\zeta_T - q_0 \theta_T)$ and $\zeta=(\zeta_T + \varepsilon_0^2 \, \theta_T /q_0)/(q_0 \gamma^2)$. This is appropriate to describe a sheared field problem, in the proximity of the rational surface, because the axisymmetric magnetic field is directed only along $\hat{\bm{\zeta}}$ at $ q_T-q_0= 0$. The gradients associated with these coordinates are: $\bm{\nabla} q_T = (q_0 s/r_0)\bm{\hat r}_T$, $\bm{\nabla} u = n_{isl} (\bm{\hat \zeta}_T - q_0 \bm{\hat \theta}_T/\varepsilon_0 )/R_0 $ and $\bm{\nabla} \zeta = (\bm{\hat \zeta}_T + \varepsilon_0 \, \bm{\hat \theta}_T / q_0)/(q_0 R_0 \gamma^2 )$. The parameter $s$ is the magnetic shear calculated at the rational surface. It is useful to rewrite the gradient of the coordinates $u$ and $\zeta$ in the form $\bm{\nabla} u = \bm{\hat u}/\rho_0 $ and $\bm{\nabla} \zeta = \bm{\hat \zeta}/Z_0$, where $\rho_0= r_0 / (q_0 n_{isl} \gamma )$, and $Z_0 = \gamma q_0 R_0$. In fact, we see that $2 \pi \rho_0$ is the length of the magnetic island in the $u$ direction and $2 \pi Z_0$ is the length of a magnetic field line of the axisymmetric equilibrium at $q_T = q_0$. In other words, $2 \pi Z_0$ is the periodicity length of the magnetic flux tube defined as the region inside the magnetic-island separatrix.

The axisymmetric magnetic field components can be written now by considering that $\hat{\bm{\zeta}}$ forms an angle $\alpha_0=\arctan (B_{pol}(q_0)/B_{tor})$ with $\hat{\bm{\zeta}}_T$. Therefore in these variables the magnetic field of an equilibrium with a magnetic island is ${\bm B}=B_{q_T,ph} \bm{\hat{r}}_T +B_{u,ph} \bm{\hat{u}}+B_{\zeta,ph} \bm{\hat{\zeta}}$, with:
% \begin{equation}
% B^{q_T}_{ph}  =  B_{isl} \sin(u) \; , \;
% B^u_{ph}  = \frac{\varepsilon_0 B_0 (q_T-q_0)}{q_0^2\gamma^2} \; , \;
% B^\zeta_{ph} =  B_0 \nonumber
% \end{equation}
\begin{equation}
B_{q_T,ph}  =  B_{isl} \sin(u) ,
B_{u,ph}  = \frac{\varepsilon_0 B_0 (q_T-q_0)}{q_0^2\gamma^2} ,
B_{\zeta,ph} =  B_0 \nonumber
\end{equation}
where we dub the components rescaled with the length of the corresponding basis vector as \emph{physical} components (see Appendix~\ref{appendix:differential-operators}).
We have adopted the \emph{constant-}$\psi$ approximation for the magnetic island, assuming a constant magnetic field amplitude of magnetic island perturbation, $B_{isl}$. The X points of the magnetic island are at $(q_T-q_0,u)=(0,0)$ and $(0,2 \pi)$  and the O point at $(0,\pi)$.
The flux surfaces of this equilibrium are labeled by $\psi= (q_T-q_0)^2 /2 + M (\cos u +1)$. M is a constant determined by the condition ${\bm\nabla}\psi \cdot \bm B = (\partial \psi / \partial q_T) (dq_T/dr_T)B_{q_T} +  (\partial \psi / \partial u) |{\bm\nabla}u| B_u$    $=0$. We obtain $M=(q_0 |s| /n_{isl}) (B_{isl}/ B_{pol,0})$, where $B_{pol,0}$ is the poloidal magnetic field calculated at the rational surface.

\begin{figure}[t]
\begin{center}
\includegraphics[width=0.47\textwidth]{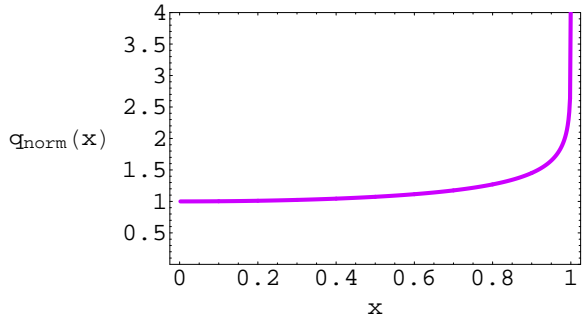}% Here is how to import EPS art
\caption{\label{fig:q(x)} The safety factor defined inside a magnetic island, plotted versus the radial position $x$, and normalized to $q (0) = 1/(\sqrt{M}n_{isl}) =\gamma / (|s|\sqrt{1-e})$. For typical magnetic islands, $1-e < 10^{-2}$ and we have $q(0) >10$.}
\end{center}
\end{figure}

\subsection{Coordinates inside the magnetic island}
\label{subsec:coordinates-inside}

Here, we model the plasma inside the magnetic island as a straight flux tube, where $\zeta$ is the translational symmetry coordinate, and $\psi$ labels the nested flux surfaces. The magnetic axis and island O point are at $\psi = 0$, while the separatrix is labeled by $\psi = \psi_{sx}= 2 M$. We define a complete set of cylinderlike coordinates $(\rho,\theta,\zeta)$, with radial-like and anglelike coordinates given by:
\begin{equation}
\rho  =  \frac{r_0}{q_0 s} \sqrt{2 \psi} \; , \;\;
\theta  = \arccos (\sqrt{M(\cos u + 1)/\psi} \,)
\end{equation}
% \begin{eqnarray}
% \rho & = & \frac{r_0}{q_0 s} \sqrt{2 \psi} \nonumber\\
% \theta & = & \arccos (\sqrt{M(\cos u + 1)/\psi} \,) \nonumber
% \end{eqnarray}
With these definitions, the magnetic axis is at $\rho=0$ and the separatrix radius is $\rho =\rho_{sx}$, which corresponds to the magnetic island  half-width $W_{isl}$, given by the Rutherford formula~\cite{Rutherford73}:
\begin{displaymath}
\frac{W_{isl}}{r_0} = 2 \sqrt{\frac{B_{isl}}{q_0 s n_{isl} B_{pol,0}}} = \frac{\rho_{sx}}{r_0} = \frac{2}{q_0 \gamma n_{isl}} \sqrt{1-e}
\end{displaymath}
with $e$ defined below in this paragraph. The angle $\theta$ is defined in the domain $(0,\pi/2)$ and extended to $(0,\pi)$ by reflection symmetry with respect to $\theta=\pi/2$, with values $0,\pi$ at the rational surface $q = q_0$. Further extension to $(0,2\pi)$ is obtained by reflection symmetry for $\rho \leftrightarrow -\rho$.
We also note that the flux surface's cross section in the $(\rho,\theta)$ plane and in the proximity of the O point is an ellipse, and the value of its  eccentricity is $e = 1 - M n_{isl}^2 \gamma^2 /s^2$. Typical magnetic islands in tokamak experiments have eccentricities close to $e\simeq 1$. The case of zero eccentricity corresponds to a value of magnetic island width, which is never encountered in tokamak plasmas because a saturation or a disruption occurs long before the island grows up to these sizes. Nevertheless, in the following sections we calculate the SAW continuous spectrum for the case $e\simeq 0$ too. In fact, this case shows clearly the analogy with the well-known results of continuous spectrum in a tokamak, whose cross section has typically small values of eccentricity.
The differential operators in this coordinate set are provided in Appendix~\ref{appendix:differential-operators}. These are necessary to give an explicit form to the equation for the SAW dynamics, that will be introduced in Sec.~\ref{subsec:model}.

\subsection{Equilibrium and safety factor}
\label{subsec:equilibrium-safety}

The equilibrium magnetic field in the region inside the island, can be described in the cylinderlike coordinates $(\rho,\theta,\zeta)$  defined in Sec.~\ref{subsec:coordinates-inside}. The contravariant \emph{physical} components are:
\begin{equation}
B^\rho_{ph} = 0   \; , \;\; B^\theta_{ph} = B^\theta_{ph,0} \sqrt{a} \, \rho \;,\;\; B^\zeta_{ph} = B_0   \nonumber
\end{equation}
Here $ B^\theta_{ph,0} =\varepsilon_0 |s| B_0 /(q_0 r_0 \gamma^2)$ is the value of $ B^\theta_{ph,0}$ at $\rho=1$, $\theta = \pi/2$, where all parameters refer to the original axisymmetric equilibrium. The function $a$ and the definition of contravariant \emph{physical} components are provided in Appendix~\ref{appendix:differential-operators}.

We define the safety factor $q$ inside the magnetic-island flux tube as the average of $ Q = \rho B_0 / ( Z_0 B^\theta ) $ over $\theta$, with $B^\theta = B^\theta_{ph} F \sqrt{1-e} / (\sqrt{a} \,\rho)$; i.e.,
\begin{equation}\label{eq:safety-factor}
q =  \frac{2}{\pi} \frac{\gamma}{|s| \sqrt{1-e}} \mathtt{K}(x) =  \frac{2}{\pi} \frac{1}{\sqrt{M} n_{isl}} \mathtt{K}(x)
\end{equation}
where we have introduced the complete elliptic integral of the first kind $\mathtt{K}(x) = \int_0^{\pi/2} d\theta /F$, with $ F= \sqrt{1-x^2 \cos^2\theta}$ and $x=\rho / \rho_{sx}$. A similar definition for the safety factor was given in Ref.~\cite{swartz84}, but our result is a factor $q_0$ smaller. The basic difference of our derivation from that of Ref.~\cite{swartz84}, is that we take into account that the length of a magnetic island flux tube is $Z_0 \simeq 2\pi q_0 R_0$, whereas in Ref.~\cite{swartz84} a flux tube with length $Z_0 \simeq 2 \pi R_0$ is considered, which is the major circumference of the tokamak (see also Ref.~\cite{Biancalani2010ppcf}).

The safety factor inside a magnetic island, normalized to the minimum value $q(x=0)$, is shown in Fig.~\ref{fig:q(x)}. We see that $q(x)$ has a singular behavior near the separatrix $x=1$, since $B^\theta$ vanishes at $\rho=\rho_{sx}$, $\theta=0,\pi$. This feature is due to the X points and is analogous to safety factor profiles in diverted tokamak plasmas.

\begin{figure}[t]
\begin{center}
\includegraphics[width=0.47\textwidth]{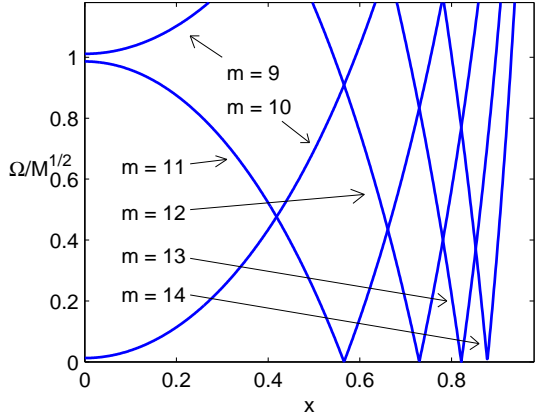}
\caption{\label{fig:cont-from-q} Continuous spectrum of SAW, $\Omega(x)$, evaluated from the safety factor profile as in Eq.~\ref{eq:cont-approx}, for the case $e \ll 1$ (corresponding to $M\simeq 1$). Modes with $n=10$, and $m=9$ to $14$ (where $m$ is the poloidal mode number) are considered, and $n_{isl}=1$ is chosen. Here the approximation of cylindrical symmetry is adopted, and therefore the continuum branches of the modes with different poloidal mode number $m$ intersect without interacting. Compare with Fig.~\ref{fig:e-zero} and Fig.~\ref{fig:M-small}, where the continuous spectrum is calculated retaining all geometrical effects due to the noncircularity of the flux surface's cross section.}
\end{center}
\end{figure}

\section{Shear Alfv\'en wave continuous spectrum}
\label{sec:continuous-spectrum}

\subsection{Model equations}
\label{subsec:model}

Here, we study the shear Alfv\'en wave dynamics in a general equilibrium, adopting the ideal MHD model. In next sections we introduce the geometry defined with the coordinates ($\rho,\theta,\zeta$), and give the explicit solution of the problem. The starting linear equations of our model are:
\begin{eqnarray} &&\varrho \frac{\partial \bm{v}}{\partial t} =-\bm\nabla P + \frac{\bm{J}}{c_L} \bm{\times B} \label{eq:momentum}\\
&&\bm{E}=-\frac{\bm{v}}{c_L}\bm{\times B} \label{eq:Ohm}\\
&&\frac{\partial P}{\partial t}= -\gamma_h P \bm{\nabla\cdot v} - \bm{v} \cdot \bm{\nabla} P \label{eq:state}
\end{eqnarray}
where $\varrho$ is the mass density, ${\bm v}$ the fluid velocity, $P$
the pressure, ${\bm J}$ the plasma current,
${\bm B}$ and ${\bm E}$ the magnetic and electric fields,
$\gamma_h$ the ratio of the specific heats and $c_L$ the speed of light. This set of equations is solved for the perturbed velocity, the perturbed magnetic field, and the perturbed pressure, expressed respectively as $\delta\bm{v} = -c_L \bm{\nabla} \phi \times \bm{B}/B^2$, $\delta \bm{B} = \bm{\nabla}\times\bm{A}$ and $\delta p$, where $\phi$ is the perturbed scalar potential and $\bm{A}$ is the perturbed vector potential. The set of equations~\ref{eq:momentum}, \ref{eq:Ohm}, \ref{eq:state} for shear Alfv\'en waves can be casted in the following linearized equation for the scalar potential:
\begin{eqnarray}
&&\bm\nabla \cdot \Big( \frac{\omega^2 }{v_A^2} \bm{\nabla}_\perp \phi  \Big) - \frac{4 \pi}{c_L} \bm{\nabla}\Big( \frac{J_\parallel}{B} \Big) \cdot  \big[ \bm{\nabla}\times (\nabla_\parallel \phi) \big]_\perp   + \nonumber\\
&& + 4\pi \Big[\Big(\bm{B }\times\frac{2\bm\kappa_s}{B^2}  \Big) \cdot \bm\nabla_\perp \Big] \Big[ \Big(\bm{B }\times\frac{\bm\nabla P}{B^2}  \Big) \cdot \bm\nabla_\perp\Big]  \phi +\nonumber \\
&& +\nabla_\parallel \nabla_\perp^2 \nabla_\parallel \phi - \frac{4\pi \gamma P}{B^2} \Big[ \Big( \bm{B}\times\frac{2\bm\kappa_s}{B}\Big) \cdot  \bm\nabla_\perp  \Big]^2 \phi = 0 \label{eq:general-modes}
\end{eqnarray}
where all quantities but the scalar potential $\phi$, are referred to the equilibrium. All curvature effects are retained here: $v_A$ is a function of space, and the geodesic curvature $\kappa_s$ allows the pressure to affect the SAW dynamics. The geodesic curvature is defined as the component of the curvature $\bm\kappa = (\bm{B}/B)\cdot\bm{\nabla}(\bm{B}/B)$ tangent to the magnetic flux surfaces. Equation~\ref{eq:general-modes} is valid for both global modes and radially localized modes. It can be solved in the coordinate set ($\rho,\theta,\zeta$), by substituting the differential operators defined in Appendix~\ref{appendix:differential-operators}.

Here, we want to consider the shear Alfv\'en wave propagation near the resonant flux surfaces where the energy is absorbed by continuum damping. Therefore, we focus on the dynamics of continuum modes, namely modes that are characterized by radial singular structures.
This allows us to drop the second and third terms in Eq.~\ref{eq:general-modes}. The linear equation for radially localized shear Alfv\'en modes in a compressible nonuniform tokamak plasma can be written in the form~\cite{chu92}:
\begin{equation}\label{eq:continuum-modes}
\frac{\omega^2}{\omega_A^2}\nabla_\perp^2\phi + Z_0^2 \nabla_\parallel \nabla_\perp^2 \nabla_\parallel \phi - \frac{\omega_{BAE-CAP}^2}{\omega_A^2} \nabla_\perp^2\phi= 0
\end{equation}
where $\omega_A = v_A/Z_0$. We adopt the value of the frequency of the low-frequency SAW continuum accumulation point (CAP), delimiting the frequency gap of the beta induced Alfv\'en eigenmode (BAE)~\cite{chu92,turnbull93}, given in Ref.~\cite{zonca96}:
\begin{equation}\label{eq:BAE}
\omega_{BAE-CAP} = \frac{1}{R_0} \sqrt{\frac{2 T_i}{m_i}\Big( \frac{7}{4} + \frac{T_e}{T_i}  \Big)}
\end{equation}
where $T_i$ and $T_e$ are the ion and electron temperatures, and $m_i$ is the ion mass.
Here, we focus on frequencies higher than $\omega_{BAE-CAP}$ and consistently neglect kinetic effects associated with wave-particle resonances~\cite{zonca96}. The operators $\nabla_{\parallel}$ and $\nabla_{\perp}$ are the gradients calculated along and perpendicularly to the equilibrium magnetic field (given by the sum of the axisymmetric field plus the island field).

\subsection{Eigenvalue problem}
\label{subsec:Eigenvalue-problem}

In section~\ref{subsec:coordinates-inside}, we have described the region inside a magnetic-island flux tube with a cylinderlike set of coordinates, labeling the magnetic flux surfaces with a radial-like coordinate $\rho$. This magnetic flux tube is assumed here to be the static equilibrium of the problem. We want to put the equation for radially localized shear Alfv\'en modes, Eq.~\ref{eq:continuum-modes}, which is written in a general vectorial form, in an explicit form, using a cylinderlike set of coordinates for a straight flux tube and the differential operators defined in Appendix~\ref{appendix:differential-operators}. We substitute the parallel and perpendicular differential operators:
\begin{eqnarray}
Z_0 \nabla_\parallel & = & \sqrt{M} n_{isl} F \frac{\partial}{\partial \theta} + \frac{\partial}{\partial \zeta} \\
\rho_{sx}^2 \nabla^2_\perp & = & a \frac{\partial^2}{\partial x^2}
\end{eqnarray}
and we impose boundary conditions of periodicity in $\theta$ and  $\zeta$ on $2\pi$ circles. Moreover, using the symmetry of the equilibrium in the $\zeta$ coordinate, we can write the general solution as $\phi = \hat\phi(x,\theta) \exp{(-i n \zeta)}$, where $n$ is the mode number in the $\zeta$-direction.

\begin{figure}[t]
\begin{center}
\includegraphics[width=0.47\textwidth]{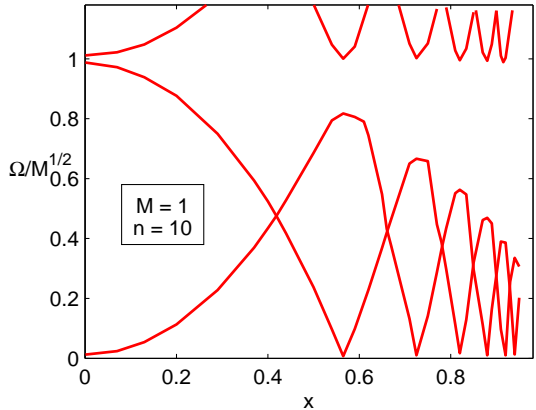}
\caption{\label{fig:e-zero} Continuous spectrum of SAW,  $\Omega(x)$, obtained by solving numerically Eq.~\ref{eq:eigenvalue}, and plotted versus the radial position inside the island, $x$. The case of small eccentricity case, $e \ll 1$ (corresponding to $M\simeq 1$), is considered. Typical values of the equilibrium parameters have been chosen and $n_{isl}=1$. The O point is at $x = 0$ and the separatrix at $x=1$. Modes with $n=10$ are considered. The MiAE gap is found at frequencies $\Omega \sim \sqrt{M} n_{isl}$.}
\end{center}
\end{figure}

We also note that the problem we are considering is greatly simplified if we change coordinates from $(\rho,\theta,\zeta)$ to a field aligned set of coordinates $(\rho,\theta,\xi)$, where $\xi = \zeta - \chi (x,\theta)$, and $\chi = (\sqrt{M} n_{isl})^{-1} \int_0^\theta d\theta' / F(x,\theta')$ is the normalized incomplete elliptic integral of the first kind. In fact, we have $\bm{B}\cdot\bm{\nabla}\xi = 0$ and the parallel derivative takes the simple form $Z_0 \bm{\nabla}_\parallel = \sqrt{M} n_{isl} F \, \partial/\partial \theta$. Finally, Eq.~\ref{eq:continuum-modes} is written in the form of an eigenvalue problem:
\begin{equation}\label{eq:eigenvalue}
\Big[ \frac{\Omega^2}{M n_{isl}^2} + \frac{F}{a} \frac{\partial}{\partial \theta} a F \frac{\partial}{\partial \theta}\Big] \tilde\phi'' = 0
\end{equation}
where $\Omega^2 = (\omega^2-\omega_{BAE-CAP}^2)/\omega_A^2$ is the eigenvalue, and $\tilde\phi'' = (\partial^2 \hat\phi / \partial^2 x) \exp{(-i n \chi)}$ is the eigenfunction. The boundary condition $\hat\phi''(\theta=0)=\hat\phi''(\theta=2\pi)$ now reads $\tilde\phi''(\theta=0)=\tilde\phi''(\theta=2\pi) \exp{(2\pi i n q)}$, with the safety factor $q$ defined in Eq.~\ref{eq:safety-factor}.

Before solving Eq.~\ref{eq:eigenvalue} in its exact form, we note here that we can have an approximate solution by assuming cylindrical symmetry, and using the safety factor as calculated in Eq.~\ref{eq:safety-factor}. In this case, the solution of Eq.~\ref{eq:eigenvalue} can be written in the form:
\begin{equation}\label{eq:cont-approx}
\Omega = (nq - m)/q 
\end{equation}
where we have used $\partial/\partial \theta = i\, m$, with $m$ being the poloidal mode number. When cylindrical symmetry is assumed, the continuum branches of modes with different $m$ are not coupled, and therefore intersect without interacting (see Fig.~\ref{fig:cont-from-q}). 
In the case of small eccentricity $e \ll 1$, and close to the O point $x\simeq 0$, we have $a =  F = 1$ and the formula given in Eq.~\ref{eq:cont-approx} is a good approximation of the solution of Eq.~\ref{eq:eigenvalue}: $\Omega = \sqrt{M} n_{isl} (nq - m)$.

In next section, we provide the exact solution of Eq.~\ref{eq:eigenvalue}, namely retaining all geometrical effects due to the noncircularity of the flux surface's cross section.

\subsection{Solution}
\label{subsec:Solution}

In previous section, the  equation for SAW continuous spectrum has been written in the form of an eigenvalue problem in the coordinates $(\rho,\theta,\xi)$, Eq.~\ref{eq:eigenvalue}. This is an ordinary differential equation in one variable, $\theta$, where the radial position $x$ is treated as a parameter. It can be solved numerically, with a shooting method code for each position $0<x<1$, giving  the continuous spectrum $\Omega^2 (x)$ as result.

The case of small eccentricity, corresponding to $M=1$ (and $e\ll 1$), is considered firstly. Typical values for the equilibrium parameters, $q_0 = 2$, $s=1$, $\varepsilon_0 = 0.1$, $n_{isl} = 1$ are chosen. Continuum modes with mode number $n=10$ are considered (see Fig.~\ref{fig:e-zero}). In this case of small eccentricity, we see that near the O point, $x\simeq 0$, cylindrical symmetry is a good approximation and the result is well described by the formula given in Eq.~\ref{eq:cont-approx}, as shown in Fig.~\ref{fig:cont-from-q}. On the other hand, for $x \sim 1$, the flux surfaces have a noncircular cross section and therefore mode coupling occurs between modes with different $m$ numbers. This mode coupling is responsible for the presence of a gap in the spectrum analogous to the EAE gap in tokamaks~\cite{betti91}, dubbed here as magnetic-island induced AE (MiAE) gap. The closer a flux surface is to the separatrix, the wider is the MiAE gap.

\begin{figure}[t]
\begin{center}
\includegraphics[width=0.47\textwidth]{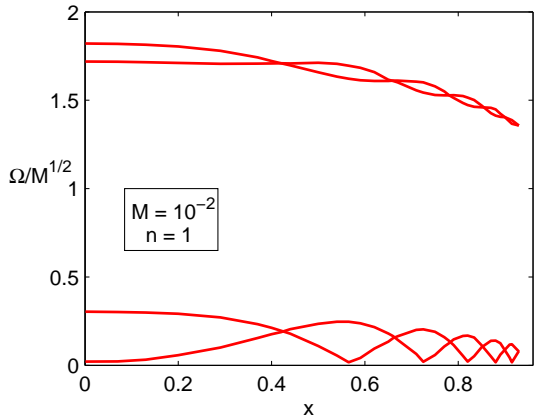}
\caption{\label{fig:M-small} Continuous spectrum of SAW, $\Omega(x)$, calculated as in Fig.~\ref{fig:e-zero}, for $n=1$ and a typical size magnetic island, $M=10^{-2}$ (corresponding to $e\simeq 0.99$). In this case, the structure of the continuous spectrum is the same as for $M=1$, but the MiAE gap is much wider. This implies that a wide range of frequencies exists, where modes contained within the magnetic island are not affected by continuum damping.}
\end{center}
\end{figure}

The calculation is also repeated for a magnetic island with a typical size, with $M = 10^{-2}$, using the same equilibrium parameters. This corresponds to a magnetic island with $W_{isl} \simeq \sqrt{1-e} \, r_0 = 0.1 \, r_0$, which is the order of magnitude of a saturated magnetic island in tokamaks.
For this case, we considered modes with $n=1$. We find that the continuous spectrum structure, shown in Fig.~\ref{fig:M-small}, is the same as for $M=1$, but in this case we obtain a much wider MiAE gap. This is due to the fact that flux surfaces of a typical magnetic island have a high eccentricity, and therefore the mode coupling between modes with different $m$ numbers is stronger. The MiAE-gap central frequency is proportional to the magnetic island half-width and near the O point we have:
\begin{equation}\label{eq:MiAE-gap}
\Omega_{MiAE} = \sqrt{M} n_{isl}  = \frac{q_0 s n_{isl}}{2} \frac{W_{isl}}{r_0}
\end{equation}
This is valid for any value of any value of magnetic island half width, corresponding to any value of eccentricity.

For typical tokamak plasma parameters, $\omega_{BAE-CAP}^2/\omega_A^2 \sim \beta q_0^2$, with $\beta$ denoting the ratio between plasma and magnetic pressures. Therefore, finite compression BAE frequency gap and magnetic island induced MiAE gap are of comparable size when $M \sim \beta q_0^2$, i.e. for a typical saturated magnetic-island size in tokamaks.
We have considered a straight flux tube model, neglecting the effect of curvature coupling among different $m$ numbers. Therefore, a gap analogous to the toroidicity induced Alfv\'en eigenmode gap~\cite{cheng85,Cheng86} is not found here. Nevertheless, such a toroidal MiAE gap is expected to be present inside magnetic islands, at frequencies lower than the frequencies of the ellipticity MiAE gap. Its gap width can be estimated by substituting $v_A$ with $v_A/(1+\varepsilon_0 \cos\theta_T)$ in Eq.~\ref{eq:continuum-modes}. This gives a multiplication factor $(1+2\varepsilon_0 \cos\theta_T)$ to the first and third terms of Eq.~\ref{eq:continuum-modes}, and consequently to the first term of Eq.~\ref{eq:eigenvalue}. Therefore, we expect to have a toroidicity induced gap within a magnetic island, with a width of the order of $\Delta \Omega^2 \simeq M n_{isl}^2 \varepsilon_0$, analogously to the theory of toroidicity induced AE. The magnetic island also has a modulation in the tokamak toroidal direction, given by the number $n_{isl}$. This helicity of the flux tube is neglected here and is expected to weakly couple modes of the SAW continuous spectrum with different $n$, analogously to the case of stellarators. These effects are not expected to qualitatively modify the MiAE gap and will be investigated in a different work as further extension of the present theory.
On the other hand, the BAE gap is considered in our model in the definition of $\Omega^2$, resulting in an upward shift in the continuous spectrum of the frequencies $\omega^2 / \omega_A^2$.

\section{Conclusions and discussion}
\label{sec:Discussion}

One of the main damping mechanisms of SAW instabilities in nonuniform plasmas is continuum damping, which occurs near the magnetic flux surfaces where the frequency of the instability matches the SAW continuum frequency.
Therefore, understanding the structure of the SAW continuous spectrum is one crucial step in studying the stability properties of a tokamak plasma.  In toroidal devices, a frequency gap in the continuous spectrum represents a range of frequency where discrete Alfv\'en eigenmodes (AE) can grow unstable, practically unaffected by continuum damping.

The radial structure of the continuous spectrum of SAW is calculated in this paper in an equilibrium inside a magnetic-island flux tube. 
Since the typical island frequency and growth rate are much lower than the SAW oscillation frequency, we have assumed that the equilibrium magnetic field is given by the usual tokamak axisymmetric field plus a quasi-static helical distortion due to the magnetic island.
The flux tube is considered to be straight, in order to focus on the effects of the noncircularity of the cross section. Curvature effects are retained only in the formation of the BAE gap in the low frequency part of the continuous spectrum. A linear MHD model is adopted for finite-beta tokamak plasmas. A generalized safety factor is defined inside the magnetic-island flux tube, giving information on the rational flux surfaces where different modes couple to create gaps in the SAW continuous spectrum. The linearized SAW equation is solved with a shooting method code inside a finite-size magnetic island and the result compared with the SAW continuous spectrum calculated in tokamak equilibria.

We have found that there exists a SAW continuous spectrum within a magnetic island, similar to that calculated in tokamak equilibria~\cite{Biancalani2010prl}.  For the case of small eccentricity of the flux surfaces, modes with different poloidal numbers are not coupled and therefore, the branches of the continuous spectrum intersect and there is no formation of gaps. This is shown to occur inside a magnetic island with very large (not typical) width, near the O point, where the flux surfaces are nearly circular. On the other hand, a typical-size magnetic island is shown to have wide gaps in the continuous spectrum, due to the strong eccentricity of the flux surfaces, labeled here as MiAE gap. This is analogous to the formation of EAE gaps in tokamaks, but is found here inside the magnetic island equilibrium. Note that MiAE can exist as bound states within the island, essentially free of continuum damping, provided that plasma equilibrium effects and free energy sources can drive and bind them locally~\cite{hasegawa74}.

The equilibrium inside a magnetic island is shown here to be analogous to a tokamak equilibrium, and, consequently, we emphasize the analogies between AE within the magnetic islands (MiAE) and the well-known AE in tokamaks. The MiAE peculiarity resides in the corresponding SAW continuous spectrum dependence on the magnetic-island size. Moreover, we expect to find all kinds of Alfv\'enic instabilities inside a magnetic island, and not only ellipticity-MiAE. For instance, toroidicity-MiAE could grow inside a gap due to the flux tube toroidicity effects, which are not studied here and will be treated in a subsequent paper.

We point out the important implications of this result in understanding the dynamics and stability properties of a magnetic island. In fact, the presence of MiAE inside a magnetic island could modify the equilibrium profiles and nonlinearly affect the magnetic island growth. On the other hand, MiAE could nonlinearly affect energetic particles redistribution in the proximity of the magnetic island rational surface, caused in part by the radial magnetic field due to the magnetic island itself~\cite{brudgam2010}.

The frequency of the ellipticity MiAE gap has been shown to be proportional to the magnetic-island width [see Eq.~\ref{eq:MiAE-gap}]. This suggests the possibility of using the MiAE frequency scalings as a novel magnetic-island diagnostics in a fashion similar to other commonly used Alfv\'en spectroscopy techniques~\cite{pinches04,Breizman05,Zonca09}. Unlike in the BAE case, the radial MiAE localization at the center of the island makes them difficult to detect by external measurements, and makes the use of internal fluctuation diagnostics, such as electron cyclotron emission and soft x rays, necessary. By detecting MiAE inside the island and measuring their frequency, one has indirect information on the magnetic-island size.

\section*{Acknowledgments}
This work was supported by the Euratom Communities under the contract of Association between EURATOM/ENEA and in part by PRIN 2006 and CREATE, as well as by DOE Grants No. DE-FG02-04ER54736 and No. DE-FC02-04ER54796, and by the NSF grant No. ATM-0335279.
Part of this work was done while one of the authors, A.B. was at the University of Pisa, in collaboration with University of California, Irvine, and Centro Ricerche ENEA Frascati, which are gratefully acknowledged for the hospitality. One of the author, A.B. would also like to thank Andreas Bierwage for valuable suggestions in writing the shooting method code. Useful discussions with C. Di Troia, I. Chavdarovski, X. Wang and P. Porazik are also acknowledged.

%\vspace{-0.4 cm}
%\section*{Bibliography}

\newpage

\appendix

\section{Differential operators}
\label{appendix:differential-operators}

%\vspace{-0.4 cm}
Here, we calculate the differential operators in the set of coordinates inside the magnetic island $(\rho,\theta,\zeta)$. The domains of these coordinates are $0 < \rho < \rho_{sx}= W_{isl}$, $0 < \theta < 2\pi$ and $0 < \zeta < 2\pi$, where $\zeta$ is the coordinate of translational symmetry, and periodicity in $\theta$ and $\zeta$ is assumed for the perturbations. The gradients of these coordinates are:
\begin{eqnarray}
\bm{\nabla} \rho & = & \sin\theta \, \hat{\bm{q}} - \cos\theta \sqrt{1-e}  F  \, \hat{\bm{u}} = \sqrt{a} \, \hat{\bm{\rho}}\nonumber \\
\bm{\nabla} \theta & = & ( \cos\theta \, \hat{\bm{q}} + \sin\theta \sqrt{1-e}  F  \, \hat{\bm{u}} ) /\rho  = \sqrt{b} \, \hat{\bm{\theta}} /\rho \nonumber \\
\bm{\nabla} \zeta & = & \hat{\bm{\zeta}}/ Z_0 \nonumber
\end{eqnarray}
where we use the notation $\hat{\bm{V}} = \bm{V}/V$ for a unitary length vector.
Moreover we have the relation $\bm{\nabla} \rho \cdot \bm{\nabla} \theta = c/\rho $. Here $F = \sqrt{1 - x^2 \cos^2 \theta}$, where $x=\rho/\rho_{sx}$, and:
\begin{eqnarray}
a & = & \sin^2\theta + \cos^2\theta (1-e)  F^2 \nonumber \\
b & = & \cos^2\theta + \sin^2\theta (1-e)  F^2 \nonumber \\
c & = & \cos\theta \sin\theta \,(1 - (1-e) F^2) \nonumber
\end{eqnarray}

A covariant basis ($\bm{g}_\rho ,\bm{g}_\theta, \bm{g}_\zeta$) can also be defined as orthogonal to the contravariant basis ($\bm{g}^\rho = \bm\nabla \rho ,\bm{g}^\theta = \bm\nabla \theta , \bm{g}^\zeta = \bm\nabla \zeta$), namely satisfying: $\bm{g}^i \cdot \bm{g}_j = \delta^i_j$. With this notation, any vector can be written as $\bm{V}= \sum_i V^i \bm{g}_i = \sum_i V^i_{ph} \hat{\bm{g}}_i$, where we call $V^i_{ph}$ the \emph{physical} components of the vector, $V^i_{ph} = V^i |\bm{g_i}|$.
Finally, the differential operators for the cylinderlike set of coordinates $(r,\theta,\zeta)$ are:
\vskip -1em
%\begin{widetext}
\begin{eqnarray}
\bm{\nabla} f & = &  \sqrt{a} \,  \frac{\partial f}{\partial \rho} \hat{\bm{\rho}} + \frac{\sqrt{b}}{\rho} \,  \frac{\partial f}{\partial \theta} \hat{\bm{\theta}} + \frac{1}{Z_0} \frac{\partial f}{\partial \zeta} \hat{\bm{\zeta}}  \\
\bm{\nabla} \cdot \bm{V} & = & \frac{1}{Z_0} \frac{\partial V^\zeta_{ph}}{\partial \zeta}+ \frac{\sqrt{1-e} \, F}{\rho} \Big(  \frac{\partial}{\partial \rho} \Big(\frac{\rho V^\rho_{ph}}{\sqrt{b}}\Big) + \frac{\partial}{\partial \theta} \Big(\frac{V^\theta_{ph}}{\sqrt{a}} \Big)  \Big) \\
\nabla^2 f & = & \frac{F}{\rho} \frac{\partial}{\partial \rho} \Big( \frac{\rho}{F} \Big(  a \frac{\partial f}{\partial \rho} +  \frac{c}{\rho} \frac{\partial f}{\partial \theta}\Big) \Big)    + \frac{F}{\rho} \frac{\partial}{\partial \theta} \Big( \frac{1}{F} \Big(  c \frac{\partial f}{\partial \rho} +  \frac{b}{\rho} \frac{\partial f}{\partial \theta}\Big) \Big) + \frac{1}{Z_0^2} \frac{\partial f}{\partial \zeta^2}  \\
(\bm{\nabla}\times \bm{V})_\rho & = & \frac{\sqrt{b}}{\rho} \frac{\partial V^z_{ph}}{\partial \theta} + \frac{1}{F \sqrt{1-e}} \Big( \frac{c}{Z_0} \frac{\partial V^\rho_{ph}}{\partial \zeta} - \frac{\sqrt{a}\sqrt{b}}{Z_0} \frac{\partial V^\theta_{ph}}{\partial \zeta} \Big)  \\
(\bm{\nabla}\times \bm{V})_\theta & = & -\sqrt{a} \frac{\partial V^z_{ph}}{\partial \rho} + \frac{1}{F \sqrt{1-e}} \Big( \frac{\sqrt{a}\sqrt{b}}{Z_0} \frac{\partial V^\rho_{ph}}{\partial \zeta} - \frac{c}{Z_0} \frac{\partial V^\theta_{ph}}{\partial \zeta} \Big)\\
(\bm{\nabla}\times \bm{V})_\zeta &=& \frac{F}{\rho} \Big( \frac{\partial}{\partial \rho} \Big( - \frac{\rho c V^\rho_{ph}}{F\sqrt{b}} + \frac{\rho \sqrt{a} V^\theta_{ph}}{F}    \Big) +  \frac{\partial}{\partial \theta} \Big( \frac{\sqrt{b} V^\rho_{ph}}{F} + \frac{c V^\theta_{ph}}{F \sqrt{a}}    \Big)
  \Big)
\end{eqnarray}
%\end{widetext}
It is straightforward to see that, in the case $e\simeq 0$ and in proximity of the O point ($F = a = b = 1$, $c=0$), these differential operators reduce to the known differential operators in cylinder geometry.

\vskip -1em

\end{document}